\documentstyle[preprint,epsfig,aps]{revtex}
\tightenlines
\begin{document}

\title{Electromechanical Limits of Polymersomes}
\draft
\author{H. Aranda-Espinoza$^{1,}$\footnote[2]{On leave: Departamento de 
F\'{\i}sica, Centro de Investigaci\'on y Estudios Avanzados, M\'exico}, 
H. Bermudez$^2$, F. S. Bates$^3$ and D. E. Discher$^{1,2}$}
\address{$^1$Institute for Medicine and Engineering and $^2$School of 
Engineering and 
Applied Science,
University of Pennsylvania, Philadelphia PA 19104, $^3$Department of Chemical 
Engineering and Materials 
Science, University of Minnesota, Minneapolis MN 55455}
\date{\today}
\maketitle
\begin{abstract}

Self-assembled membranes of amphiphilic diblock copolymers enable comparisons of 
cohesiveness with lipid
membranes over the range of hydrophobic thicknesses $d=3-15$ $nm$. At zero mechanical 
tension 
the breakdown potential $V_c$
for polymersomes with $d = 15$ $ nm$ is $9$ $ V$ compared to $1$ $ V$ for liposomes with 
$d = 3$ $nm$. Nonetheless,
electromechanical stresses at breakdown universally exhibit a $V_c^2$ dependence, and 
membrane capacitance
shows the expected strong $d$-dependence, conforming to simple thermodynamic models. The 
viscous nature
of the diblock membranes is apparent in the protracted post-poration dynamics.

\end{abstract}
\pacs{PACS Numbers: 82.70.Uv, 87.68.+z, 68.65.-k}

\begin{figure}
\epsfxsize=3in
\centerline{\epsfbox{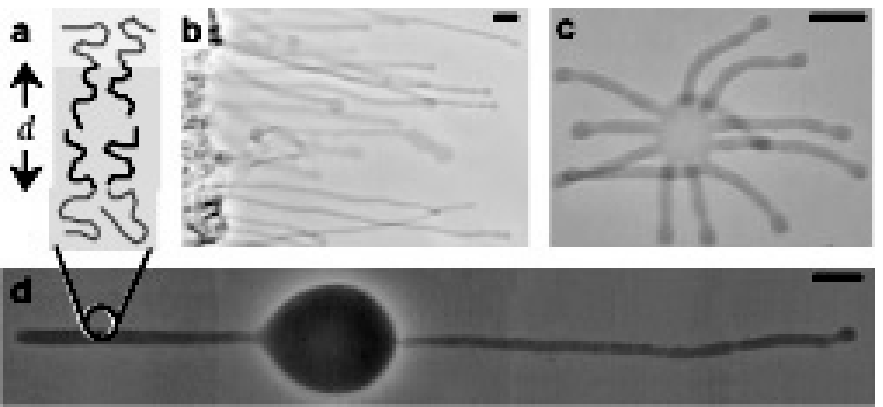}}
\caption{(a) Polymersome membrane schematic. (b) Vesicular tubes sprouting out from bulk OB18 
copolymer. (c) and (d)
Vesicles of OE7 copolymer exhibiting tube-like arms. Vesicles are observed using phase 
contrast with isotonic
solutions of sucrose (inside) and glucose (outside). Scale bars are $10$ $\mu m$.}
\label{fig1}
\end{figure}

A primary task for any biological membrane is to separate inside from out, partitioning 
ions and other
solutes that generate a transmembrane potential $V_m$. Electrically excitable cells, 
particularly
neurons, control and propagate this potential for purposes of signaling and 
inter-communication
\cite{alberts}.  Efforts to exploit and better understand such phenomena have most recently 
motivated
the semi-patterned growth of nerve cells on circuits \cite{neurons} as well as the 
generation of
artificial
networks with soft nanotubes pulled from vesicles \cite{networks}.  Further opportunities in
such directions now seem possible with wholly synthetic block copolymers that, analogous to 
lipids in
water, self-assemble into membranes, minimizing interfacial exposure of
hydrophobic segments (Fig.~\ref{fig1})
and thereby generating vesicles termed polymersomes \cite{discherfamily}.

Physical limits of self-assembled lipid membranes impose important constraints on 
electrochemical signals. 
Indeed, the operating range of biomembrane excitations, such as the action potential of a 
neuron, is
generally less than $0.1$ $V$ over time scales of milliseconds.  At the same time, a finite 
resting tension
$\tau$ is exerted on most cell membranes, including neurons \cite{hochmuth}, through
substrate
adhesion and additional cellular mechanisms.  From optical trap techniques and other methods, 
$\tau$ is
now generally believed to be in the range of $0.01$ to $0.1$ $ pN/nm$ \cite{dai}, though 
additional transient
stresses on cells can readily magnify such tensions \cite{simson}.
For example, electrocompressive stresses
arise with a voltage drop across a membrane and have long been recognized as coupled, most 
simply, to
lateral tensions through an integrated form of the Lippmann equation \cite{lippmann}:

\begin{equation}
\tau + \frac{1}{2} C_m V^2_m = \tau _{net}
\label{lippmanneq}
\end{equation}

\noindent
where $C_m$ is the capacitance per unit area of the membrane, and $\tau_{net}$ is an equivalent 
tension.
Alternative descriptions of electromechanical coupling have been put forward and include a thermodynamic 
study of block copolymer lamellae perforation that was specialized in its comparisons to lipid membranes 
\cite{netz}.  Direct tests of lipid membrane cohesion by Needham and Hochmuth \cite{needham} have clearly 
demonstrated the electromechanical limits of lipid membranes via the combination of micropipette 
aspiration and electroporation techniques.  From these and related tests it has become clear that 
transmembrane potentials higher than $1$ $ V$ and mechanical tensions
higher than $\sim 10$ $ pN/nm$ are not 
sustainable in lipid membranes \cite{hochmuth,needham}. Clearly, a hydrophobic thickness of only 
$d = 3 - 4$ $ nm$ for such membranes is a determinant of electromechanical stability.

Formation of polymersomes in dilute aqueous solutions has been described elsewhere \cite{discherfamily}. 
Generally, the phase behavior of amphiphiles composed of strongly segregating segments is determined by 
the hydrophilic fraction $f$ and mean molecular weight $\bar{M}_n$ \cite{matsen}. Lipids provide initial 
guidance in the synthesis \cite{hillmyer} of biomimetic super-amphiphiles; Table~\ref{table1} shows 
$f\approx 29-39\%$ 
for vesicle-forming phospholipids and
diblock copolymers, although the latter are much higher in $\bar{M}_n$ 
and somewhat more polydisperse.  For the copolymers used here, the hydrophilic segment of polyethylene 
oxide (PEO) forms a highly hydrated brush \cite{wonbrush} while aggregation is driven by a hydrophobic 
block of either polyethylethylene (PEE) or its unsaturated homologue polybutadiene (PBD).  
As will be reported 
elsewhere \cite{bermudez}, scaling of $d$ with $\bar{M}_n$  (as well as
elastic 
properties) is consistent with the hydrophobic core behaving as a fluid-like melt.

When small pieces of our copolymers are added to water, vesicular tubes spontaneously sprout and grow 
(Fig.~\ref{fig1}(b)). Vesicle formation processes - kinetics, yields, and morphologies - exhibit temperature 
dependences that likely stem from hydration-facilitated melting of PEO \cite{hillmyer}. The vesicles 
formed exhibit a wide variety of shapes that range from multi-armed stars to hundreds of micron-long 
axon-like tubes (Fig.~\ref{fig1}) as well as flaccid spheroids. Transformations between morphologies 
are easy to achieve through osmotic adjustments of the external solution \cite{discherfamily} and thus 
reveal the semi-permeable nature of the copolymer membranes. Our first generation polymersomes composed 
of a PEO-PEE diblock copolymer (designated OE7 in Table~\ref{table1}) possess a core thickness $d \approx 8 \pm 1$ $nm$ 
and have already been shown to be mechanically tough \cite{discherfamily}.  In this Letter, we elaborate 
the more general electromechanical responses of polymersomes including two new covalently-crosslinkable 
PEO-PBD copolymers of similar block fraction but higher molecular weight (OB16 and OB18 in 
Table~\ref{table1}).  
Specific comparisons are made with our own measurements of the highly representative lipid 
stearoyl-oleoyl-phosphatidylcholine (SOPC), which 
are comparable to the results reported elsewhere \cite{needham,olbrich}. Thus, for the first time we can 
thoroughly study 
the cohesiveness of membranes as a function of their hydrophobic core thickness over the range $3-15$ $nm$.

\begin{figure}
\epsfxsize=2.8in
\centerline{\epsfbox{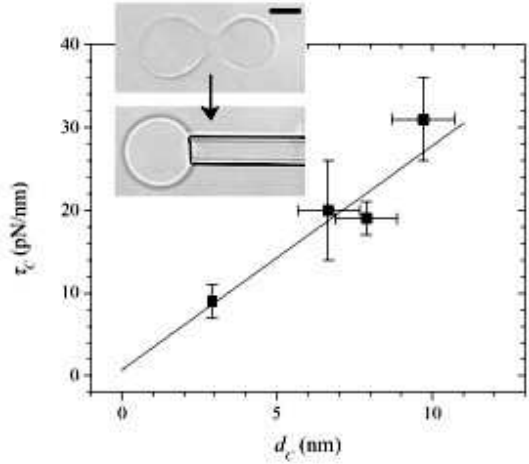}}
\caption{The lysis tension $\tau _c$ at zero applied electric field is plotted as a function of the hydrophobic
thickness $d_c$ for (left to right) SOPC, OE7, OB16, and OB18 vesicles.  Tension in the membrane is determined
from the aspiration pressure $\Delta P$ by the application of the Law of Laplace to give
$\tau = 0.5\Delta P R_v R_p /(R_v - R_p )$ where $R_v$ is the external vesicle radius and $R_p$ is the internal
radius of the micropipette.}
\label{fig2}
\end{figure}

Flaccid polymersomes were progressively aspirated into a micropipette (Fig.~\ref{fig2} inset) at stress rates of 
$\sim 0.1$ $pN nm^{-1}sec^{-1}$ up to the point of rupture. Values of the lysis tension $\tau _c$ for SOPC 
as well as the three polymersomes studied are shown in Table~\ref{table2}.  The results generally show that 
increasing $\bar{M}_n$ leads to an increase in stability, consistent with general ideas of meso-phase stability 
for strongly segregated copolymers \cite{matsen}.  As such, the hydrophobic core is expected to behave as 
a dense fluid.  Assuming incompressibility, the thickness at rupture is given by $d_c = d/(1+\alpha _c )$, 
where $\alpha _c$ is the area strain at rupture as measured directly from images of aspirated vesicles. For 
lipids, $\alpha _c$  is very small, typically $0.03 - 0.05$ \cite{olbrich}, so that $d_c \approx d$. For 
polymersomes, however, $\alpha _c$ ranges from $0.2 \pm 0.06$ for OE7 to $0.44 \pm 0.08$ for OB18. Thus, 
polymersomes 
should thin considerably.  A simple linear fit of $\tau _c$ vs. $d_c$ is found to intersect the $y$-axis 
at essentially zero tension 
(Fig.~\ref{fig2}).  Moreover, the slope of this fit defines a lysis stress 
$\Sigma _c = \tau _c / d_c \approx 27$ $atm$.
Such a stress is typical of cavitation pressures for homogeneous fluids 
\cite{cavitation} suggesting that lysis of membranes can occur through nucleation of
water-filled cavities inside the hydrophobic core. 

\begin{figure}
\epsfxsize=3in
\centerline{\epsfbox{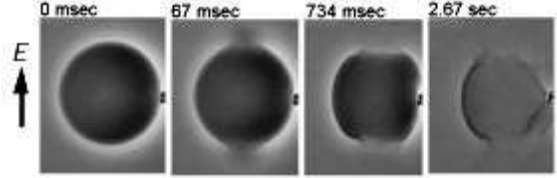}}
\caption{Snapshots of electrocompressive rupture of an OB16 polymersome with an initial 
diameter of $\sim 40$ $\mu m$.
The holding tension is below $2$ $pN/nm$, and the breakdown potential is $\sim 4$ $V$.}
\label{fig3}
\end{figure}

Electromechanical experiments were performed to determine the breakdown potential $V_c$ at different 
membrane tensions. Once again a flaccid vesicle was aspirated to a prescribed, subcritical tension 
while being 
manipulated to a position between two platinum wire electrodes separated by $\sim 1$ $mm$.  A $60$ $\mu sec$ 
square-shaped pulse was applied across the electrodes at intervals that varied from seconds to minutes 
depending on the post-poration dynamics of the vesicle (see below). The applied potential was increased at 
discrete intervals until membrane rupture was observed.
A typical electroporation event at low applied tension is shown in Fig.~\ref{fig3}. The electrodes were arranged 
parallel to the pipette, generating an electric field ${\bf E}$ as illustrated, and the suction 
pressure was sufficient 
to hold the vesicle in the micropipette. A first image taken at zero applied field ($0$ $ msec$) demonstrates 
the integrity of the membrane by showing a phase-dense vesicle interior (sucrose solution) against a 
phase-light exterior (equi-osmolar glucose).  Dramatic rupture within 1-2 video frames of the applied pulse 
was invariably indicated by a 'jet' of released sucrose;  such jets always occurred at focal point(s) on 
the membrane where the surface normal is parallel/antiparallel to ${\bf E}$.  However, 
the dynamics of pore formation differed considerably from one membrane system to another.  Membranes 
composed of OB16 (Fig.~\ref{fig3}) typically showed {\it two} very large, antipodal pores that grew to many microns in 
size over seconds.  Continued growth often completely rendered the OB16 polymersomes, despite the low tensions. 
Phospholipid vesicles, in contrast, showed more rapid and even reversible pore dynamics at vanishing tension, 
as described by others \cite{zhelev,sandre}.  Polymersomes of the smallest copolymer, OE7, exhibited 
liposome-like dynamics whereas polymersomes composed of the largest copolymer, OB18, were far more 
protracted in their poration dynamics. With OB18, encapsulated sucrose was invariably lost over periods of 
{\it tens} of seconds, with weak, barely visible jets implying relatively small but sustained pores.

Kinetic diversity in membrane poration is understood at a first level in terms of the interplay between edge 
energy, characterized by a line tension $\lambda$ that tends to close the pore, and the work done by any 
{\it dynamic} membrane tensions that tend to force the pore open. A balance between these two energies 
gives the metastable pore size $r_0 = \lambda / \tau $ above which the pore will grow - provided sufficient 
tension is maintained.  Sandre {\it et al}. \cite{sandre} recognized the important role of viscosities in 
vesicle pore dynamics.  In the limit of vanishing tension, the rate of pore closure was 
postulated to scale as $\lambda / \eta _s $  where $\eta _s$ is the surface viscosity. Although $\lambda$ 
is expected - from simple bending energy considerations of a hydrophilic pore - to increase in linear 
proportion to $d$, separate measurements of lateral diffusion of copolymers (Lee {\it et al}. (unpublished))
suggest that $\eta _s$ increases strongly with $\bar{M}_n$ and is at least ten-fold greater in the 
present polymersomes compared to 
lipid membranes.  Moreover, the $\bar{M}_n$ of both OB16 and OB18 exceed the relevant entanglement molecular 
weights ($\sim 2$ to $3$ $kDa$), so that the polymeric nature of the present systems provides 
a clear basis for the slowed pore dynamics.

\begin{figure}
\epsfxsize=2.8in
\centerline{\epsfbox{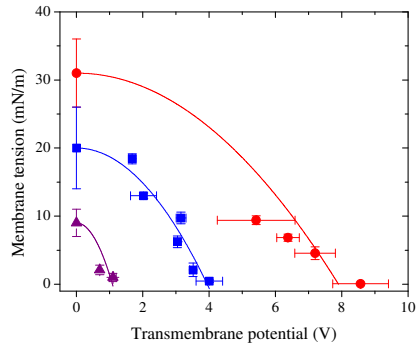}}
\caption{Membrane tension $\tau$ versus transmembrane potential $V_m$.  Rupture results are plotted for SOPC
(triangles), OE7 (squares) and OB18 (circles).  Solid lines correspond to parabolic fits using
Eq.~(\ref{lippmanneq})
in each case, giving the capacitance at rupture. Results for OB16 are omitted for clarity as 
they overlap with results for OE7.}
\label{fig4}
\end{figure}

The transmembrane potential $V_m$ has been calculated previously for a spherical shell \cite{cole} as 
$V_m = 1.5 R_v E \cos \theta [1 - \exp (-t / t_c )]$ where $R_v$ is the radius of the 
vesicle and $\theta$ is the angle between the membrane surface normal and ${\bf E}$. 
The charging time $t_c$ is given by 
$t_c = R_v C_m (\rho _i + 0.5 \rho _e)$, where $\rho _i$ and $\rho _e$ are internal 
and external 
solution resistivities, respectively. In this work, $t_c$ is orders of magnitude smaller than the pulse 
duration used.  To assess coupling of the mechanical and electrical stresses as suggested by 
Eq.~(\ref{lippmanneq}), polymersomes are aspirated to a set tension and subsequently subjected to electrical 
stresses as explained above.  At each imposed tension, $V_c$ is determined and the resulting points are 
plotted in the $\tau - V_m$ plane (Fig.~\ref{fig4}).  The data show that the higher the membrane tension, 
the lower the breakdown potential.  For OB18, the results are remarkable: this $\sim 15$ $nm$ thick 
membrane can 
transiently withstand the potential of a $9$ $V$ battery.

Fitting the rupture data ($\tau , V_c$) for each self-assembled membrane system by Eq.~(\ref{lippmanneq}) 
readily generates an amphiphile-specific estimate of $C_m$ at rupture (Table~\ref{table2}).  Within this phase 
space spanned by $\tau$ and $V_m$, the area under each parabolic segment provides a phenomenological 
measure of the range of electromechanical function or robustness.  For SOPC, this robustness is very 
small (see Table~\ref{table2}); for polymersomes, in contrast, the robustness can be orders of 
magnitude larger.  
Fitted values of polymersome $C_m$ are nonetheless very low compared to those reported for 
lipid membranes, $\sim 1$ $ \mu F / cm^2$ \cite{cole}.  This is qualitatively consistent with the 
enhanced thickness of polymersome membranes since $C_m \approx \epsilon / d_c$, where $\epsilon$ is the 
hydrophobic dielectric constant.  On the other hand, $V_c$ at $\tau = 0$ increases considerably with $d_c$ 
(see Eq.~(\ref{lippmanneq}) and Fig.~\ref{fig4}).  By definition, the surface charge at rupture $\sigma _c = C_m V_c$ 
should then be independent of $d_c$.  Indeed, Table~\ref{table2} indicates that $\sigma_c$ varies by only a factor 
of about two within this structurally diverse set of membrane systems studied.

Polymersomes made from OB16 and OB18 offer the further possibility of crosslinking the membrane core. 
Free radical polymerization in solution proves highly effective \cite{wonmicelles} with dramatic 
increases in the robustness of the membranes.  Simple osmotic rupture tests - where large vesicles are 
observed to burst after suspension in a sufficiently hypotonic solution - indicate that $\tau _c$ of 
cross-linked membranes are of order $\sim 10^3$ $pN/nm$.  Thus $\Sigma _c \approx 1000$ $atm$. 
Using such values 
in Eq.~(\ref{lippmanneq}) and assuming values for $C_m$ listed in Table~\ref{table2}, $V_c$ for the cross-linked OB's 
are estimated to exceed $125$ $ V$ for $60$ $ \mu sec$ pulses.

Electrically excitable cells, particularly neurons, control and propagate electrochemical potentials 
for signaling purposes \cite{alberts}, but the transmembrane voltages employed are always constrained 
by physical limits of a self-assembled lipid membrane system.  The results here with a series of 
self-assembling biomimetic diblock copolymers considerably expand these limits, and clearly indicate 
the overall electromechanical robustness of hyperthick membranes.  As might be expected, $\tau _c$ 
and $V_c$ both increase with $d$ while $C_m$ decreases and $\sigma _c$ hardly changes. Though a 
microscopic theoretical treatment of membrane electroporation is lacking, the capacitance and 
insulation properties of such self-assemblies provide a more physically reliable basis for 
microelectronics applications such as sensor platforms and those that might exploit conducting 
copolymer segments. In addition, drug delivery applications could benefit through formulations of 
robust copolymer vesicles that reversibly reseal or not when challenged by sufficient membrane tensions 
and/or voltages.  

\vskip.3cm
\noindent
ACKNOWLEDGMENTS HAE and HB thank R. Parthasarathy for discussions and comments. HAE was
supported by NIH/NHLBI T32HL-07954 training grant. DED 
gratefully acknowledges 
discussions with D.A. Hammer and E.A. Evans. Funding was provided by
NSF-MRSEC at the U. of Minnesota and U. of Pennsylvania, and NASA.


%

\vskip0.5in

\vfill\eject

\vskip0.5in

%
\begin{table}
\caption{Vesicle-forming amphiphiles: comparisons between lipids, SOPC and dimyristoyl-phosphatidylcholine
(DMPC) \protect\cite{olbrich}, and diblock
copolymers including
$d$ and $f$ of the formed membranes.
Cryo-TEM was used to estimate $d$ for the polymersomes \protect\cite{bermudez}.}
\begin{tabular}{dddddd}
Molecule&Average&$f$&$\bar{M}_n$&$\bar{M}_w / \bar{M}_n$&$d$ \\
 &Structure& &($kDa$)& &$(nm)$\\
\tableline
DMPC&--&0.36&0.68&1&2.5$\pm$0.06\\
SOPC&--&0.31&0.79&1&3.0$\pm$0.06\\
OE7&$EO_{40}-EE_{37}$&0.39&3.9&1.10&8.0$\pm$1.0\\
OB16&$EO_{51}-BD_{55}$&0.37&5.2&1.10&10.6$\pm$1.0\\
OB18&$EO_{82}-BD_{126}$&0.29&10.4&1.10&14.8$\pm$1.0\\
\end{tabular}
\label{table1}
\end{table}

\vfill\eject

\begin{table}
\caption{Average values for the critical tension (at $V_m = 0$) and critical voltage (at $\tau = 0$).  Fitted
values for capacitance at rupture, associated surface charge and robustness $R$.}
\begin{tabular}{dddddd}
Molecule&$\tau_c$&$V_c$&$C_m$&$\sigma_c$&$R$\cr
 &($pN/nm$)&($V$)&($\mu F/cm^2$)&($C/m^2$)&$(V pN/nm$)\\
\tableline
SOPC&9&1.1&1.5&0.016&9.9\\
OE7&20&4.0&0.26&0.01&79.7\\
OB16&19&3.95&0.24&0.096&74.8\\
OB18&31&8.6 &0.08&0.007&265.8\\
\end{tabular}
\label{table2}
\end{table}


\begin{references}
\bibitem{alberts} B. Alberts {\it et al}., {\it Essential Cell Biology} (Garland Publishing, Inc. New York, 1997).

\bibitem{neurons} P. Fromherz and A. Stett, Phys. Rev. Lett. {\bf 75}, 1670 (1995); M.P. Maher {\it et al}., 
J. Neurosci. Meth. {\bf 87}, 45 (1999).

\bibitem{networks} E.A. Evans {\it et al}., Science {\bf 273}, 933 (1996); A. Karlsson {\it et al}., Nature 
{\bf 409}, 150 (2001).

\bibitem{discherfamily} B.M. Discher {\it et al}., Science {\bf 284}, 1143 (1999); B.M. Discher {\it et al}., Curr. Opin. Coll. 
$\&$ Interface Sci. {\bf 5}, 125 (2000).

\bibitem{hochmuth} R.M. Hochmuth {\it et al}., Biophys. J. {\bf 70}, 358 (1996).

\bibitem{dai} J. Dai and M.P. Sheetz, Biophys. J. {\bf 77}, 3363 (1999); E. Evans and A. Yeung, Biophys. J. {\bf 56}, 151 (1989)

\bibitem{simson} R. Simson {\it et al}., Biophys. J. {\bf 74}, 514 (1998);  H. Ra {\it et al}., J. Cell Sci. {\bf 12}, 1425 (1999).

\bibitem{lippmann} G. Lippmann, Ann. Phys. {\bf 149}, 546 (1873)

\bibitem{netz} R.R. Netz and M. Schick, Phys. Rev. E {\bf 53}, 3875 (1996).

\bibitem{needham} D. Needham and R.M. Hochmuth, Biophys. J. {\bf 55}, 1001 (1989).

\bibitem{matsen} M.W. Matsen and M. Schick, Curr. Opin. Coll. $\&$ Interface Sci. {\bf 1}, 329 (1996); 
D.A. Hajduk {\it et al}., J. Phys. Chem. B {\bf 102}, 4269 (1998)

\bibitem{hillmyer} M.A. Hillmyer and F.S. Bates, Macromolecules {\bf 29}, 6994 (1996).

\bibitem{wonbrush} Y-Y.Won {\it et al}., J. Phys. Chem. B {\bf 104}, 7134 (2000).

\bibitem{bermudez} H. Bermudez {\it et al}., (unpublished).

\bibitem{olbrich} K. Olbrich {\it et al}., Biophys. J. {\bf 79}, 321 (2000); W. Rawicz {\it et al}., 
{\it ibid.} {\bf 79}, 328 (2000). Note that we focus on pure phospholipids; 
cholesterol is known to stiffen and thereby toughen membranes \cite{needham}, but cholesterol 
does not, on its own, 
self-assemble into a membrane. In contrast, the elasticity of all the pure amphiphilic 
self-assemblies here differ by less than $\sim$ two-fold \cite{bermudez}.

\bibitem{cavitation} F.R. Young, {\it Cavitation} (McGraw-Hill, New York, 1989).

\bibitem{zhelev} D. Zhelev and D. Needham, Biochim. Biophys. Acta {\bf 1147}, 89 (1993).

\bibitem{sandre} O. Sandre {\it et al}., Proc. Nat. Acad. Sci. {\bf 96}, 10591 (1999).

\bibitem{cole} K.S. Cole, {\it Membrane, Ions and Impulses} (University of California Press, California, 1972).

\bibitem{wonmicelles} Y-Y. Won {\it et al}., Science {\bf 283}, 960 (1999).




\end{references}
\end{document}